\documentclass[prl,twocolumn,amsfonts]{revtex4}
\usepackage{amsfonts}
\usepackage{amsmath}
\usepackage{amssymb}
\usepackage{graphics,graphicx}
\usepackage{psfrag}

\def\etal{{\it et al.\/}}

\newcommand{\be}{\begin{equation}}
\newcommand{\ee}{\end{equation}}
\newcommand{\bea}{\begin{eqnarray}}
\newcommand{\eea}{\end{eqnarray}}

\newcommand{\nn}{\nonumber}

\begin{document}

\title{Detection of acceleration radiation in a Bose-Einstein condensate}

\author{A. Retzker$^1$, J. I. Cirac$^2$, M. B. Plenio$^1$, B. Reznik$^3$}

\affiliation{$^{1}$  Institute for Mathematical Sciences, Imperial
College London, SW7 2PE, UK} \affiliation{QOLS, The Blackett
Laboratory, Imperial College London, Prince Consort Rd., SW7 2BW,
UK} \affiliation{$^{2}$ Max-Planck-Institut f¨ur Quantenoptik,
Hans-Kopfermann-Str. 1, 85748 Garching, Germany. }
\affiliation{$^{3}$  Department of Physics and Astronomy, Tel-Aviv
University, Tel Aviv 69978, Israel}

\begin{abstract}
We propose and study methods for detecting the Unruh effect in a
Bose-Einstein condensate. The Bogoliubov vacuum of a Bose-Einstein
condensate is used here to simulate a scalar field-theory, and
accelerated atom dots or optical lattices as means for detecting
phonon radiation due to acceleration effects. We study Unruh's
effect for linear acceleration and circular acceleration. In
particular, we study the dispersive effects of the Bogoliubov
spectrum on the ideal case of exact thermalization. Our results
suggest that Unruh's acceleration radiation can be tested using
current accessible experimental methods.
\end{abstract}
\maketitle

\date{\today}

One of the surprising fundamental consequences of relativistic
quantum field theory is the dependence of the concept of particle
number on the observer's state of motion. While inertial observers
see the vacuum as empty, non-inertial observers generally perceive
this vacuum as populated with particles. Unruh \cite{unruh1976}
showed that a uniformly accelerated particle detector perceive the
field in vacuum as a thermal state with temperature $k_B T_U = \hbar
a/2\pi c$, where $a$ is the proper acceleration.
The Unruh effect is related to other particle creation effects in
curved space-time, such as Hawking radiation, and the
Gibbons-Hawking thermalization in a cosmological expansion
\cite{Birrellbook}.

Numerous experimental ideas for detecting the effect have been
suggested. They include, accelerated electrons in circular high
energy accelerators\cite{Bell1983}, circular motion of electrons in
a Paul trap\cite{Rogers1988}, intense laser induced electron
acceleration\cite{Chen1999} and passage of atoms through a
cavity\cite{Scully2006}. Other setups simulate the Gibbon-Hawking
cosmological expansion thermalization effect in an expanding
Bose-Einstein condensate (BEC)\cite{Fedichev2003},  and in an
expanding linear ion trap\cite{Alsing2005}. (See also
\cite{Rosu2001}).

In this letter we propose to simulate and detect the Unruh effect
using accelerated atom dots (AD) \cite{quantum-dot1} or using optical
lattices in a BEC. Since the relevant velocity is the speed of
sound, $c_s\approx 1[mm/sec]$, $T_U\approx 10 [nK \, sec^2 /m]
\times a [m/sec^2]$ and the currently feasible acceleration of
optical lattices may reach  $a \approx 5\times10^5 [m/sec^2]$, the
Unruh temperature can be significantly higher than the relevant
energy scales, the AD minimal energy gap ($\approx 100 Hz
\approx nK $), and the BEC temperature.

Let us begin by recalling some features of the Unruh effect.
A detector is modeled as a localized system with internal levels
$|g\rangle$ and $|e\rangle=\sigma^+|g\rangle$ and energy gap $\omega_d$,
which moves along a trajectory $x_D(\tau)$ and $t(\tau)$,
where $\tau$ is the detector's  proper time. In the simplest case,
a free scalar field $\phi$, initially in its vacuum state,
 couples with the detector through
\be \label{unruh} H_{i}= g\biggl( e^{i \omega_d\tau }\sigma_+ +
e^{-i\omega_d\tau}\sigma_- \biggr) \phi(x_{D}(\tau),t(\tau)). \ee By
evaluating the transition amplitudes between the levels, it is then
found that for inertial trajectories the detector remains unexcited,
while for uniformly accelerated trajectories the detector becomes
thermalized. This can be seen by evaluating to the lowest order in
$g$ the transition amplitudes. Inserting   $x_D(\tau)={c^2\over a}
\cosh{a\tau\over c}$, and $t(\tau) ={c\over a}\sinh{a\tau\over c}$,
and the expression for a free field $\phi(x,t)$ in Eq.
(\ref{unruh}), one finds that a field mode $\omega$ has a time
dependent coupling of the form: $g_{e}(\tau,\omega)=\exp({i {\omega
c\over a}e^{-{a\tau\over c}}})$. This readily yields transition
probabilities which satisfy $P_{\rm excitation}/P_{de-excitation}=
e^{-E/k_BT_U}$,  where $T_U$ is the Unruh temperature.

It is important to note that: \break \noindent {\it i.} The
appearance of the effective coupling $g_{e}(\tau,\omega)$ is
sufficient in order to thermalize the detector. A similar coupling
is also a landmark of the Hawking and cosmological thermalization
effects. \break \noindent {\it ii.} In the Unruh effect property
{\it i.} is a direct consequence of the detector's {\em accelerated
motion}. This can be easily seen \cite{Alsing2004} by noticing that
the field mode $\omega$ is Doppler shifted in the detector's rest
frame to
 $\omega'(\tau)= \omega_0{1-v/c\over \sqrt{1-(v/c)^2}}= \omega_0 e^{-a\tau/c}$.
Therefore, the relevant collected phase factor becomes $\exp(i\int
\omega(\tau)d\tau)) =g_e(\omega,\tau)$. \break \noindent {\it iii.}
The Unruh effect is manifestly relativistic. Hence the interaction
(\ref{unruh}) is defined in the detector's rest frame, and the
trajectory, $x_{DR}(t) =c \sqrt{ t^2+ {c^2/a^2}}$ coincides with
non-relativistic acceleration only for sufficiently short times.


The above points quantify, with increasing refinement, important
aspects of the Unruh effect, which one wishes to simulate in a
specific model. For example,  {\it i.} can be obtained by modifying
the vacuum normal mode frequencies $\omega$ to $\omega(t)=\omega
e^{-{at\over c}}$, and realized in an ion traps by changing the trap
frequency \cite{Alsing2005}, or by an an expanding  BEC
\cite{Fedichev2003}. In what follows we suggest a model that
incorporates properties {\it i.} and {\it ii.}, and finally shortly
discuss possible realizations of {\it iii.}.

It is well known that small perturbations of the BEC Schr\"odinger
field satisfy a relativistic-like Klein-Gordon equation with the
speed of sound $c_s$ playing the role of $c$\cite{Garay2000}.
Nevertheless, the transformation laws for a moving detector will
remain non-relativistic. We can therefore obtain the effective
coupling constant ({\em i.}) as a consequence of non-relativistic
Doppler shift by choosing a modified trajectory: $ x_{Deff}(t)=
\left(c_s t+ {c_s^2\over a} e^{-at/c_s}\right) $ which differs from
the relativistic trajectory $x_{DR}(t)$ above (when $c=c_s$), by
$O[a^2t^3/c_s]$ for short times, and  $O[c_s^2/a^2t]$ for long
times. The Doppler shift  $\omega'= \omega_0(1-v/c_s)= \omega_0
e^{-at/c_s} $, has the same time dependence as in the relativistic
case, with $\tau\to t$.  We hence expect that a suitable detector
that moves along $x_{Deff}$ will be similarly thermalized.

Consider then a setup with hyperfine levels, $a$ and $d$, where $a$
forms a condensate  described by the  field $\Psi$. Level $d$ will
be used for an AD produced by a localized potential
$V_d$\cite{Diener,quantum-dot1} or by an optical lattice. It will be
sufficient to consider only one level with a wavefuction $\psi_d(x)$
and creation and destruction operators $d$, $d^\dagger$.
Since $V_d$ affects only atoms in the state $d$, in the absence of
further coupling with the condensate, moving about $V_d$ will not
disturb the condensate state. We need however to make sure that
nonadiabatic excitations are negligible.
The adiabatic condition in this case can be derived by transforming
to the AD rest frame and for the trajectory $x_{Deff}(t)$ is
given by: $v T = a^2 / (\omega^2 c_s) \ll x_0$ where $x_0$ is the
width of the wave function. For trap frequency $\omega\approx
100kHz$, $a/c_s\approx\omega_d\approx 100Hz\approx 10^{-3}
\omega$, and since the l.h.s of the inequality is less than $1 \AA$, the
condition is  satisfied. Atomic levels then couple
through elastic collisions, which to the lowest order redefin the
detuning $\delta$, and produce self interaction terms $g_{dd}
d^\dagger d^\dagger d d$. A large $g_{dd}$ is used
\cite{quantum-dot1} to simulate a two-level detector (Eq.
\ref{unruh}). In the following we found more convenient to assume
small $g_{dd}$, hence the detector is a harmonic oscillator.

We couple between the AD and the BEC by laser induced Raman
transitions described by interaction Hamiltonian \be
\label{atom-dot} H_{int}=  \delta d^\dagger d+ \Omega_a \int dx\psi_d(x) \biggl(
d^\dagger \Psi(x) + {\rm h.c.}\biggr),
 \ee
where $\Omega_a$ is the Rabi frequency.
 At first sight Eq.
(\ref{atom-dot}) lacks the number non-conserving terms of Eq.
(\ref{unruh}), which are essential to the effect. However our
interest is in the resulting coupling with phonons.  Using
Bogoliubov's theory we expand the field operator
\be \hat\Psi(x)= \phi(x) + \sum_k
u_k(x) e^{-i\omega t} c_{k} + v_k(x) e^{+i\omega t} c_{-k}^\dagger
\label{bogoliubov}
 \ee
where $\phi(x)$ is a c-number,  and $u_k(x),v_k(x)$ and $c_k$ are the phonon
mode functions and annihilation operators. This brings the BEC
Hamiltonian to a free field form $ H_{BEC} = \sum_{k} \omega_k
c_{k}^\dagger c_{k} $, and spectrum $\omega_k= \sqrt{(c_s k)^2 +
\big({k^2\over 2m})^2}$ that is ``relativistic", $\omega\approx k$,
for $k<k_c= m c_s/\hbar$.

Inserting Eq. (\ref{bogoliubov}) into Eq. (\ref{atom-dot}), and
assuming that $\psi_d(x)$ extends over scales smaller
then the phonon wavelength,
(the dominant coupling arises from long wavelengths), we obtain \bea
&H_{int}& = \delta d^\dagger d+ \sqrt{n_a}\Omega_a(d+ d^\dagger)+ \\
&& \Omega d^\dagger \sum_k( u_k(x_D)e^{-i\omega_k t}c_k + v_k(x_D)e^{i\omega_k
t}c_{-k}^\dagger) + h.c., \nn\eea
where  $n_a$ is the effective number of condensate atoms at the AD.
For $k\ll k_c$, $u_k\approx v_k$,
this model coincides with Unruh's detector model Eq. (1),
apart from the term $\sqrt{n_a}\Omega_a(d+ d^\dagger)$ which describes
the interaction with the mean-field. This term can be eliminated using a two mode condensate with levels $a$ and $b$ that couple as in (Eq. 2)
via Raman transitions and with Rabi frequencies satisfying $\Omega_a=-\Omega_b$. Cancelation of this term is then obtained from the symmetry of the Hamiltonian.
Alternatively, one can use a single mode condensate and remove the displacement in the AD final state by applying the unitary $\exp({\sqrt{n_a}\Omega\over 2\delta^2}(d-d^\dagger))$. This approach requires a precise control of $\sqrt{n_a}$\cite{correct}.

Consider the effect of $H_{int}$ on the AD when the
condensate is in its ground state: $c_{k,\alpha}|BEC\rangle=0.$
 For a uniform motion $x=vt$,  the excitation amplitude is to first order
$ i\int_{-T}^T dt \sum_k v_k(x_D(t)) e^{i(\omega_d +
(1-v)\omega_k))t} \overrightarrow{\scriptstyle{T \rightarrow
\infty}} \sum_k\delta(\omega_d+(1-v/c_s)\omega_k)$. Therefore as
long as $v<c_s$, the detector remains  unexcited. For the suggested
non-inertial trajectory $x_{Deff}$, as long as $k<k_c$, $v_k(x) \sim
u_k(x) \sim \exp(ikx)$, the transition amplitudes reduce to
$A_{\pm}(\omega)\propto \int^T_{-T}exp (\pm i\omega_db t- i\omega_k
e^{-at}) dt$, which coincides with Unruh's expressions, with $t$
replacing $\tau$.
 The total transition probability, is then $P_{\pm} = \sum_k |A_\pm(k)|^2$.
For a given mode $k$ the contribution to $A_\pm(k)$ comes from the
saddle point at $t\sim t_s(\omega)\equiv {c_s\over a}\log
{\omega/\omega_d} \pm {1\over\sqrt{\omega_d a}}$; for longer
interaction times, in order to recover the Unruh effect larger
momenta are required, with $k$ that grows exponentially with the
duration of the interaction. Consequently, in a realistic situation
the finiteness of $k_c$ will cause deviations for acceleration time
$T>t_s$.  To resolve this difficulty we have to restrict the
interaction time and study in more detail the deviations due to the
$k$ dependence of ${v_k \over u_k}$, and the dispersion relation.

In a experiment of finite time $T$, it is
important to consider with care the modification due to the temporal
change of the coupling strength. In the simplest case of abrupt
change in the coupling, the transition probabilities can be
approximated as  $ P_\pm(\omega) \approx\left\vert A_\pm(\omega)
+\frac{1}{i\omega_d}e^{\mp i\omega_d t_s(\omega)}\right\vert^2.$
Since different modes contribute at different times, the total
contribution is effectively averaged and  $
P_\pm\propto\vert A_\pm \vert^2+ \frac{1}{\omega_d^2}.$ The
correction does not decrease with the energy gaps since $\vert
A_\pm\vert^2$ scale as $\frac{1}{\omega_d a}$.
More generally we shall  assume that the coupling starts and ends
smoothly over a time scale $\gamma$ by adding a regulator
$e^{-t/\gamma}$, i.e., a slow decoupling function.

In the following we have assumed that the detector is accelerated for time
$T$ and that the experiment is repeated $n$ times by moving the AD  back and forth in the BEC. We have studied numerically a BEC
with a finite number of $N$ phonon modes and described the detector
by a harmonic oscillator. The total state is then described by a covariance matrix,
and detector's population and temperature are derived from the
 AD reduced covariance matrix.
 \begin{figure}
\psfrag{a}{$\gamma$} \psfrag{w}{$\omega_d$} \psfrag{(a)}{(a)}
\psfrag{(b)}{(b)}
\begin{center}
\includegraphics[width=0.40\textwidth,height=0.20\textheight]{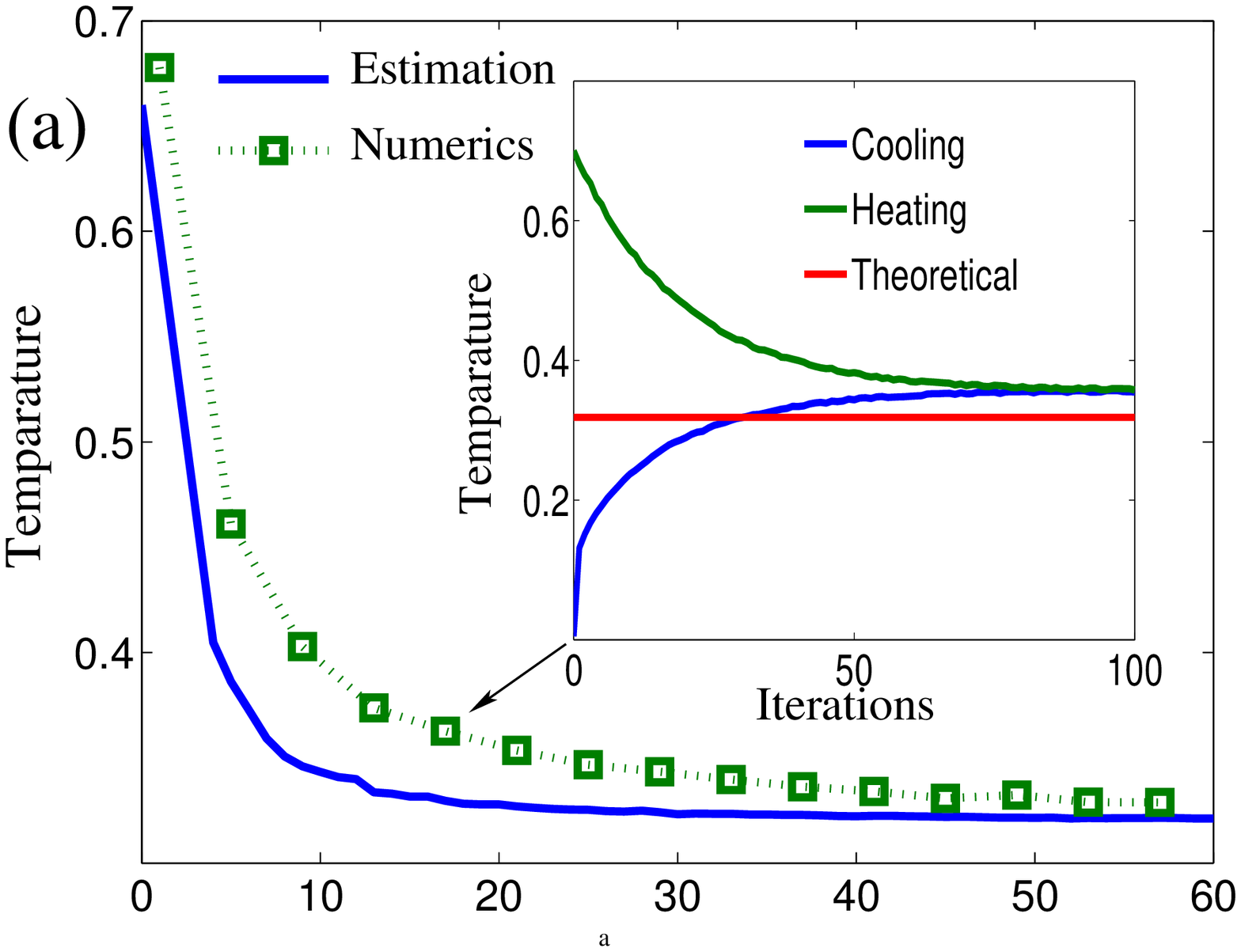}
\includegraphics[width=0.40\textwidth,height=0.10\textheight]{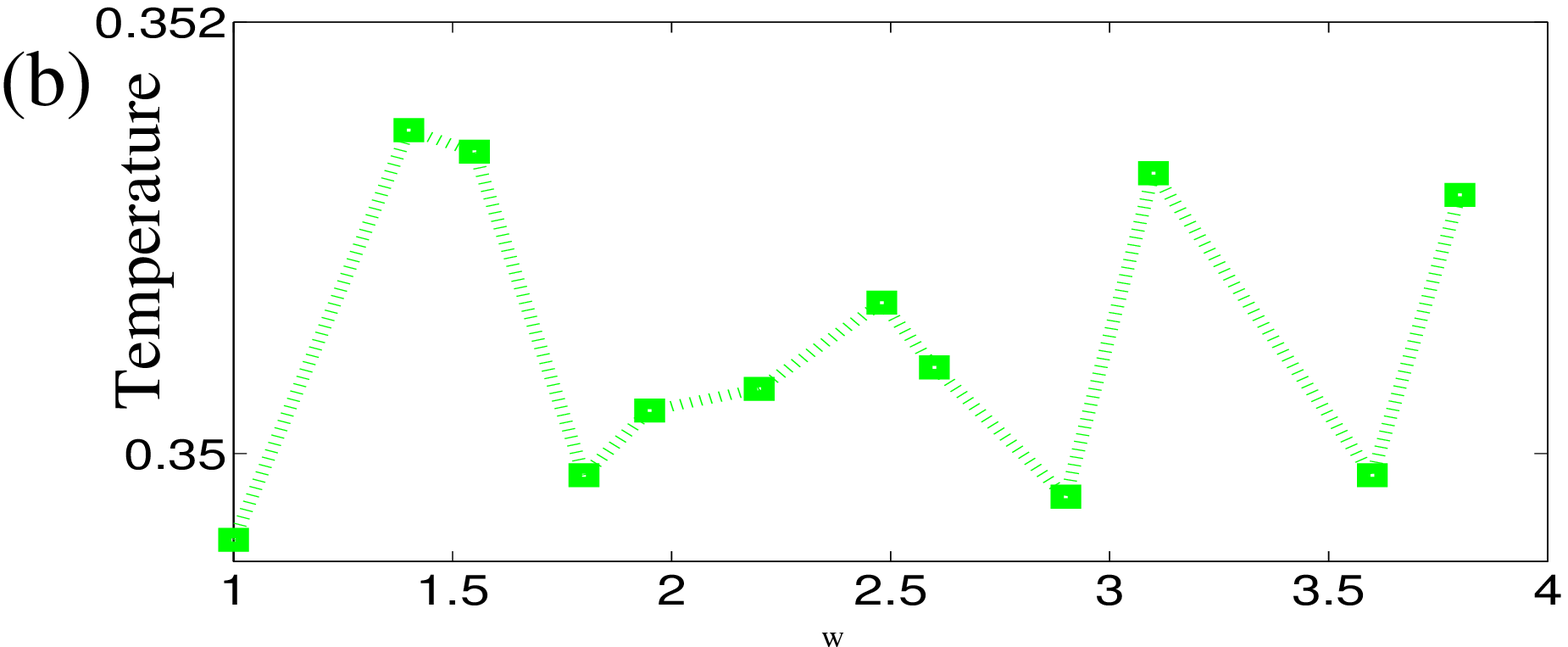}
\end{center}
\caption{1(a)Thermalization of an oscillator coupled of the smearing
time $\gamma$. The field is modeled by $20$ modes,$a=2,\omega_d=1$
and the coupling is $1/50$.
 The Green line denotes the numerical results.
The Blue line is derived by numerical integration of the transition
amplitudes. The inset shows the thermalization curve for
$\gamma=17$. The red line shows the theoretical temperature. The
blue and green graphs shows the cooling and the heating curves. The
results are in the units of $\omega_d$. 1(b) In order to check
whether the distribution is thermal we changed the energy gap and
calculated the temperature. }\label{ThermAlpha}
\end{figure}

We first considered the ideal case with $k_c\to \infty$. As is shown
in Fig. \ref{ThermAlpha}(a) the effective temperature of the
detector changes gradually until it reaches a final steady state
after $n \sim 100 $ repetitions. The temperature is slightly higher
then the value of $T_U$ since the finite decoupling time and the
final coupling strength increas the average final energy of the
steady state. By increasing $\gamma$ we can get closer to the
theoretical value of $T_U$. In Fig. \ref{ThermAlpha}(b) the final
temperature is plotted for various values of the detector energy
gaps. As can be seen the temperature remains unchanged in agreement
with a thermal distribution, up to fluctuations of $\Delta T/T\sim
1\%$. The corrections observed here is due to the finiteness of the
number of modes and the interaction time.
\begin{figure}
\begin{center}
\psfrag{T}{Time[$2\pi/\omega_d$]}
\includegraphics[width=0.4\textwidth,height=0.25\textheight]{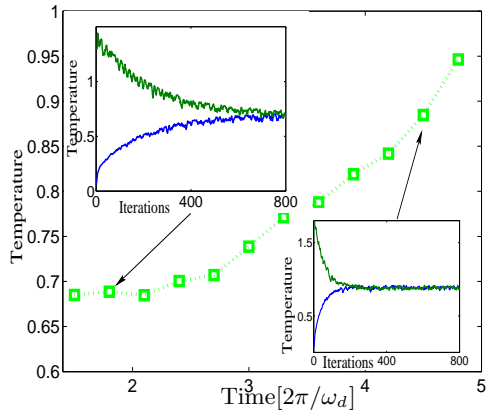}
\end{center}
\caption{The final temperature using the full Boguliubov theory, as
a function of the interaction time $t_0$ (with $\gamma=0$). The
ratio of the energy gap and the cutoff energy was taken as $1:500$,
hence deviation are expected for $\omega\ge120 \omega_d$. This
corresponding to acceleration times
  $t_0\approx 1/a \log 120\approx 2.4$. Indeed, the theoretically predicted temperature, $T\approx
0.67$, is obtained in this simulation for $t<t_0$.
}\label{tc}
\end{figure}

Next we extended the analysis to the full problem with a finite
cutoff scale $\omega_c=c_s k_c$ which corresponds in a realistic BEC
to more than $10kHz$, and is two orders of magnitude larger than the
atomdot's minimal energy gap, which is limited by the fluctuations
of the laser. There are two types of corrections. The first type is
due to the changing dispersion relation; since the phase in the
transition amplitude is now given by, $e^{\pm i\omega
_{d}t}e^{i(kx-\omega t)}=e^{\pm i\omega
_{d}t}e^{i(ckt-cke^{-at}/a-\omega (k)t)}=e^{\pm i\omega
_{d}t}e^{i(\left( ck-\omega (k)\right) t-cke^{-at}/a)}$, the
detector's energy gap is corrected by $ck-\omega (k)$, which is
always a negative quantity. For certain modes the effective detector
gap can vanish, which implies a divergence in the resulting partial
excitation probability. For higher modes the temperature can then
becomes negative, which causes a gradual population inversion since
$P_{\pm}=\frac{2\pi c_s }{\pm\omega _{d}a} \frac{1}{e^{\pm 2\pi
\omega _{d} c/a}-1},$ the ratio for large frequency tends to unity.
This cutoff effect would be felt once $\omega_d=ck-\omega(k)$, which
is smaller than the field cutoff. The second type of correction
comes from the modified momentum dependence of mode functions $u_k$
and $v_k$. As $k$ increases $v_k$ decreases to zero, hence for
$T>t_c(k_c)$  the temperature starts decreasing.


In order to observe the Unruh effect, we can reduce the effects of
the above `ultra-high' frequency corrections by selecting a
sufficiently short time scale. Fig. \ref{tc} displays the resulting
final temperature for a numerical computation which includes all
Boguliubov's theory corrections.  The expected thermalization effect
can be observed but due to the shorter interaction time requires a
slightly higher number of repetitions, $n\sim 300$. In order to
decrease the number of repetitions the initial state can be chosen
at the vicinity of the final temperature. To avoid finite
temperature corrections, we need to have the phonon number in the
relevant interacting modes to be smaller then 1. For a BEC
temperature of $50 nk$ this requires a gap energy on the order of
$1kHz$.
\begin{figure}
\begin{center}
\psfrag{time}{Time[$2\pi/\omega_d$]}
\includegraphics[width=0.40\textwidth,height=0.15\textheight]{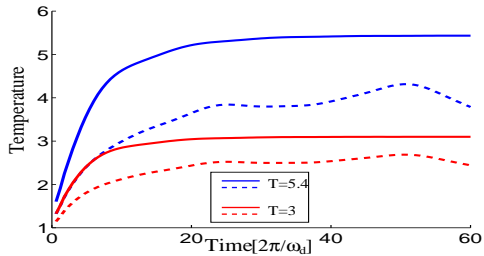}
\end{center}
\caption{Simulation of thermalization and finite size effects in circular motion.
The final analytically computed temperatures is $3[\hbar \omega_d]$ (red lines) and $5.5[\hbar \omega_d]$
(blue lines).  To examine finite size effects we compare between
two cases: the upper graphs for which the system size $L>c_s t$ ($L/c_s =125[2\pi/\omega_d]$)
and boundary effects are minor, and
the lower graphs  for which $L<c_st$ ($L/c_s =25[2\pi/\omega_d]$)
and oscillations due to finite size effects become noticeable
}\label{circ1}
\end{figure}

Another interesting experimental possibility is to simulate the
effect of circular acceleration \cite{Bell1983,Levin1993}. Unlike
the ideal Unruh effect, here the accelerating detector sees the
vacuum as excited but usually is {\em not} thermalized, i.e., it's
final temperature depends on the energy gap. The advantage of this
setup is that the detector does not have to satisfy relativistic
equations and thus no special path is needed. Moreover, in the limit
where the frequency of rotation is much smaller than the energy gap,
the interaction is effectively non-zero only with a finite band of
frequencies. Consequently the effect can be insensitive to the
cutoff. The limit of $v \approx c$ is especially interesting since
the temperature divergence and the detector becomes thermalized,
making this regime ideal for experiment. The rotation of the
detector could be realized either using dipole
traps\cite{beugnon2007} or optical lattices\cite{tung2006}, in both
setups the speed of rotation could reach the speed of sound. This
makes the circular variant simpler to manifest than the linear one.
Fig.\ref{circ1} displays the numeric results for a thermalization
effect of a circulating atomic quantum dot.



We  remark that a fuller relativistic-like realization,
can be done as follows: we consider the condensate coupled to the AD as in Eq. (2),
but choose the detuning $\delta=0$.
Using Boguliubov's
expansion we then obtain the Hamiltonian $H_r\approx\Omega(t)[
\phi(d+d^\dagger) + (d-d^\dagger)\sum_k u_k(b_k-b_k^\dagger)]\equiv
\Omega H'$, where $k<k_c$ was assumed. The first term represents the
free detector Hamiltonian (energy levels have become superpositions
of number states) and the second term the interaction with the
field. The idea is then to use the common factor $\Omega(t)$ and
modify the laser intensity so that $\Omega(t)\propto {d\tau\over
dt}= {1\over \sqrt{t^2+ c_s^2/a^2}}$. Upon integration $\int H_rdt
=\int H'd\tau$,  hence this recovers the Unruh effect for a uniform-like
accelerating trajectory.


In conclusion, we found that a moving AD or an atomic lattice in a condensate
can be used to detect acceleration radiation effects that are analogous to
the Unruh effect. Our results indicate that the measurability of such effects is
within reach of current methods.
We hope that the analogy that we are making may be also useful the other way around;
that is to interpret what happens when one moves a particle in a condensate with some acceleration.



A. R. and M.B. P. acknowledge support of the European Commission
under the Integrated Project (QAP), the Royal Society and EPSRC
QIP-IRC. A. R. thanks M. khudaverdyan for many useful discussions.
B. R. and J.I. C. acknowledge support by GIF Grant no. I-857.


\begin{thebibliography}{10}
\bibitem{correct} The initial vacuum state is also effectively displaced,
however within the relevant parameters this effect can be ignored.
\end{thebibliography}
\end{document}